# A Retrofittable Photoelectron Gun Proposal for Improved Low Voltage Imaging

*Frances Quigley[1,2*], Clive Downing[2], Cormac McGuinness[1], Lewys Jones[1,2]*

*1. School of Physics, Trinity College Dublin, Dublin 2, Ireland*

*2. Advanced Microscopy Laboratory, Centre for Research on Adaptive Nanostructures & Nanodevices (CRANN), Dublin 2, Ireland*

## Abstract

Low voltage scanning electron microscopy is a powerful tool for examining surface features and imaging beam sensitive materials. Improving resolution during low voltage imaging is then an important area of development. Decreasing the effect of chromatic aberration is one solution to improving the resolution and can be achieved by reducing the energy spread of the electron source. Our approach involves retrofitting a light source onto a thermionic Lanthanum Hexaboride ($LaB_6$) electron gun as a cost-effective low energy spread photoelectron emitter. The energy spread of the emitter's photoelectrons is predicted to be between 0.11eV-0.38eV depending on the photon energy of the UV light source. Proof of principle images have been recorded using this retrofitted photoelectron gun and an analysis of its performance is presented.

*Corresponding Author: fquigley@tcd.ie

## Introduction

Across the scientific disciplines low voltage scanning electron microscopy (<5keV) has a range of applications. In the semiconductor industry the decrease in knock on damage in samples when imaged at lower voltages is extremely advantageous (Pawley, 1984). In addition to this, at low voltages the electron-specimen interaction is concentrated near the surface of the sample, this ensures signals are generated from only a shallow depth and enables the user to examine surface features at a much greater level of detail (Liu, 2000). Non-conducting samples also may benefit from low voltage imaging, where if the voltage is chosen correctly high-resolution surface features can be observed without the effects of charging (Joy & Joy, 1996). Biological laboratories have benefited from this charge neutrality in imaging uncoated soft materials (Liang et al., 2021).

Despite this wealth of applications for low voltage imaging, its use can be limited by the effect of chromatic aberration on the image resolution (Pawley, 1984). Chromatic aberration is the premature focus of lower energy spread electrons on the optic access in comparison to higher energy spread electrons and can be described by the radius of the chromatic defocus blur ($r_{chr}$), given as:

$$r_{chr} = C_c \frac{\Delta E}{E} \alpha, \qquad\qquad 1$$

where $C_c$, $\alpha$, E, and $\Delta E$ are the chromatic aberration coefficient, the beam convergence semi-angle, the beam energy and the energy spread of the electrons respectively (Goodhew et al., 2000). As this radius is inversely proportional to the energy of the electron beam, the lower the acceleration voltage of the microscope the more dominant the effect of chromatic defocus blur becomes. There are two main approaches to decreasing the effect of chromatic defocus blur. The first involves decreasing the $C_c$ coefficient using $C_c$ -correctors. These however are not widely available and can be extremely costly (Joy, 2005). As the chromatic defocus blur is proportional to the energy spread of the electrons, reducing this energy spread is the second main solution to decreasing the effects of chromatic aberration on the resolution of the image.

Decreasing the energy spread of an electron source in a SEM/Transmission Electron Microscope (TEM) usually involves the upgrade from a large energy spread electron source such as a thermionic electron gun (~1-3 eV) (Haine & Cosslett, 1961) to a low energy spread electron gun such as a Schottky field emission gun (FEG) (~0.7eV) (Tuggle et al., 1985) or cold FEG (~0.3eV) (Goldstein et al., 2003) (Williams & Carter, 2009). These lower energy spread electron guns are naturally a more costly upgrade for the user due to the more complex nature of FEGs and their requirement for higher vacuum conditions (Quigley et al., 2022).

This paper describes a new type of cost effective retrofittable electron source which utilises the photoelectric effect with the potential to produce low energy spread electrons. We will discuss the effect of key variables of photon energy and light source power on the emitter's performance and show example experimental images.

## Background

The photoelectric effect occurs when sufficiently energetic photons stimulate the direct release of electrons from near the surface of a material. Figure 1 a) and b) is after previous work by Sawa (Sawa et al., 2017), these figures show the Fermi distributions of a material struck by two light sources with different photon energies ($E_A$ and $E_B$ respectively). It describes how when the energy of the photon is greater than the workfunction ($\Phi$) of the sample, electrons will be emitted. Choosing a material with a low workfunction is therefore beneficial for a photoelectron gun as it would not require a light source with an excessively high photon energy to stimulate photoelectron production.

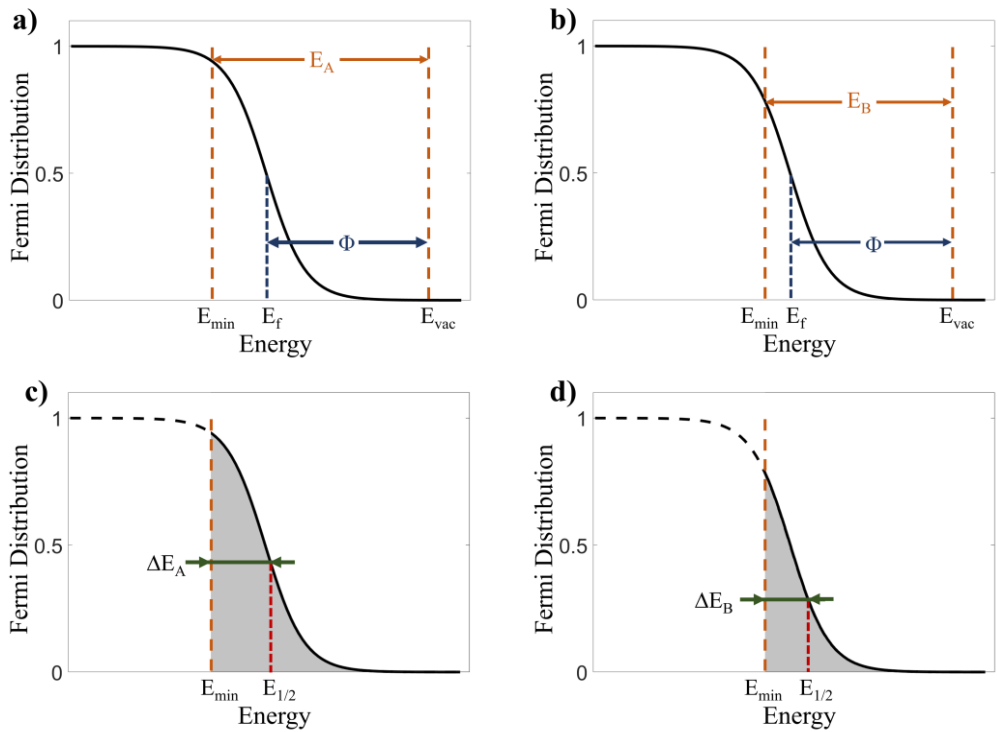

*Figure 1 a), b) Fermi distribution showing the minimum energy level ($E_{min}$) a photon of energy $E_A$ and $E_B$ can excite an electron respectively. Also shown is the workfunction ($\Phi$) which is equal to the vacuum level ($E_{vac}$) minus the Fermi level ($E_f$) of LaB$_6$. c), d) The Fermi distribution showing the energy spread ($\Delta E_A$ and $\Delta E_B$ respectively) of the emitted photoelectrons between the green arrows where $E_{1/2}$ satisfies the equation $f(E1/2)=f(Emin)/2$ (Sawa et al., 2017).*

Lanthanum Hexaboride ($LaB_6$) is a low workfunction material that is commonly found in thermionic electron guns, it has already been well investigated for its use as a photocathode in literature (Mogren & Reifenberger, 1987; May et al., 1990; Oettinger, 1990; Qian et al., 1995; Leblond et al., 1996) and is a strong candidate for a photoelectron source. Cerium hexaboride ($CeB_6$ or CeBix) is another very similar material (Torgasin et al., 2017) offered by some filament manufacturers. Other photoelectron sources utilising negative electron affinity (NEA) photocathodes have previously been investigated for low voltage SEM imaging (Ohshima & Kudo, 2004). These sources have been found to achieve brightnesses as high as a Schottky FEG at an energy spread estimated to be less than 0.2eV (Morishita et al., 2020). While a $LaB_6$ photocathode may potentially be unable to reach these levels of brightness, these NEA photocathodes usually require extra space in the vacuum chamber for surface preparation devices to clean the cathode, need illumination ports to introduce the laser light and usually operate in an extremely high vacuum chamber (Ohshima, 2003). Our work is focussed on using low cost, off the shelf components where possible, and is designed to be retrofittable on a conventional SEM. Therefore, a commercially available $LaB_6$ crystal which can be easily cleaned via resistive heating was chosen for this retrofittable photoelectron gun.

In choosing a suitable light source for the photoelectron gun, Figure 1 c) and d) must be considered. These graphs show the same Fermi distribution with the same two light sources as Figure 1 a,b) respectively but concentrate on the energy-spread of the electrons emitted. As the light source in Figure 1 a,c), has a larger photon energy than the light source in Figure 1 b,d) (i.e. $E_A > E_B$), the deeper below the Fermi level the electrons are able to be stimulated from. This is demonstrated by the shaded area under the graph in grey which represents the distribution of electrons eligible to be emitted as photoelectrons. This area implies that the larger the energy of the photon inducing photoemission, the larger the energy spread of the electrons emitted (i.e. $\Delta E_A > \Delta E_B$). Experimental evidence has already been shown that by using photon energy close to the value of the workfunction of the sample you can produce electrons with an energy spread as low as 0.11 eV (Sawa et al., 2017), around one third of the energy spread of a cold FEG (Goldstein et al., 2003). It is interesting to note that Figure 1 c) and d) show a probability versus energy distributions, analogous to a zero-loss peak that will be familiar to TEM spectroscopists.

The shaded area in these figures is of course not itself a measure of current, as that depends on the flux of photons.

When aiming to construct a practical emitter with a usable electron beam-current, the quantum efficiency ($\eta$) of LaB$_6$ is an important variable to consider when deciding on the light source, as it can be used to determine what variables influence the photoelectron current ($I_{PE}$). $\eta$ can be expressed as

$$\eta = \frac{I_{PE}/e}{P/E_{ph}} \qquad\qquad 2$$

where $P$ is the power of the light source, $E_{ph}$ is the incident photon energy and $e$ is the elementary electric charge (Sawa et al., 2017). We have plotted this value of $\eta$ as a function of energy drawn from the literature of various types of light sources (see supplemental Figure 1). Equation 2 can be rearranged in terms of I$_{PE}$ to give

$$I_{ph} = \frac{\eta \cdot P \cdot e}{E_{ph}}. \qquad\qquad 3$$

This rearrangement illustrates how the higher the power of your light source the larger the photoelectron current, as we might expect. This means that a high-power light source with monochromatic photon energy close to the workfunction of LaB$_6$ would be an appropriate choice for our photoelectron source. The survey of literature in supplemental Figure 1 indicates that many of the high photon energy light sources which produce a high quantum efficiency are gas or pulsed lasers systems which are not the most cost effective. When choosing a light source with photon energy close to the work function of LaB$_6$, LEDs and laser diodes were the main choice in literature. LEDs tend to have a large angular (~120 degrees), and wavelength spread (~$\pm$10 nm) relative to the laser diodes (~21 degrees and ~$\pm$5 nm).

Based on these considerations continuous wave laser diodes of wavelength 405nm +/- 5nm (3.06eV $\pm$ 0.038eV), which are commonly used in a variety of applications from laser engravers to curing resin in 3D printers, were chosen as the light source. Our laser packages have an additional focussing lens to make their optical path more parallel, however it is not possible to measure the final angular distribution intensity of the light as it would have travelled through a series of optical elements before striking the LaB$_6$ crystal.

Sawa et al used the following equation to theoretically predict the energy spread ($\Delta E_e$) of their electron source;

$$\Delta E_e = k_B T \, ln\left[2 \exp\left(-\frac{E_{ph}-\phi}{k_B T}\right) + 1\right] + E_{ph} - \phi \qquad 4$$

where E$_{ph}$ is the photon energy, k$_B$ is the Boltzmann constant and T and ɸ are the temperature and workfunction of the LaB$_6$ crystal (Sawa et al., 2017). Using equation 4, the theoretical energy spread of our electron source with a LaB$_6$ cathode with ɸ=2.69 which is struck by a light source of E$_{ph}$=3.06eV ± 0.038eV is ΔE=0.37±0.04 eV. The design of the photoelectron emitter with this light source and photocathode will be described and an analysis of its performance will be investigated.

## Methodology

A ZEISS EVO SEM was used as a testbed for this photoelectron emitter prototype. A LaB$_6$ crystal in the form of a standard commercial filament (Kimball Physics LaB$_6$ cathode, model ES-423E, style 90-20) was installed into its wehnelt. The wehnelt was adapted such that two slits were cut into it and two hypodermic needles were placed within these slits. Ultra-high vacuum compatible optical fibres sourced from Thor labs were chosen to guide the light through the vacuum chamber. The MV11L1 fibres were made of differently doped silica core and cladding with a polyimide coating. They had a NA of 0.22 NA, a 100 ± 3 μm, 120 ± 3 μm and 140 ± 4 μm diameter core, cladding and coating respectively, and were compatible with wavelengths between 250nm-1200nm. They could be fed through the hypodermic needles which had an inner diameter 0.159mm and outer diameter 0.312mm. The needles and ends of the fibres were positioned such that they were ~1.6mm and ~0.7mm away from the LaB$_6$ tip respectively. Assuming the maximum angle that the light can exit the fibre is the same as the maximum incidence angle of 12.71° the light from the fibre should strike the LaB$_6$ tip as they were positioned.

Solidworks was used to design our set up as can be seen in Figure 2 of the supplemental information. A schematic of our set up which shows the laser aligned with a multimode collimator connected to a fibre optic and a fibre optic feedthrough flange is shown in Figure 2a). Inside the vacuum side of the feedthrough flange, the vacuum compatible optical fibres are fed up into the gun chamber through the pumping tree. Figure 2b) displays the light injection set up while Figure 2c) shows the set up installed in a light tight box.

COMSOL modelling of the electrostatics of this adapted wehnelt with and without the needles and optical fibres in the vacuum chamber was undertaken. The metal needles which contained the optical fibres were in contact with the wehnelt and assumed to be at the same potential. It was found that the adapted wehnelt with the needles and fibres have no detrimental effect on the trajectory of the emitted electrons. There remains a possibility that the glass fibres may experience some charging, because of our geometry, this may lead to a small astigmatism, however this could not be decoupled from other sources of astigmatism corrected by the operator during imaging. These simulations can be found in Figure 3 in the supplemental information. While two fibres were installed in the setup, one fibre seemed to be delivering a superior amount of light to the crystal indicated by a larger amount of photocurrent being detected. Due to this reason only one laser diode and one fibre was used for the course of these experiments as shown by the one fibre and light injection set up in Figure 2 a,b).

Three continuous wave laser diodes of wavelength 405nm +/- 5nm (3.06eV $\pm$ 0.038eV) of powers 200mW, 400mW and 800mW were used for the following experiments. The 400mW and 200mW laser diodes were in 16mm diameter cylindrical casings as can be seen in Figure 2b) and Figure 2a) respectively. The light injection set up in Figure 2b) consists of two precision kinematic mounts to hold the laser diode and multimode collimator such that they were aligned on an optical rail. In this set up an optical fibre FG200UEA from Thor labs with a 0.22 NA, 200 µm $\pm$ 4 µm Core which accepts wavelengths between 250 - 1200 nm was used to connect the multimode collimator to the flange with the fibre optic feedthrough. Its core, cladding and coating were made from pure silica, fluorine-doped silica, and acrylate respectively. While its cladding and coating diameter were 220 $\pm$ 2 µm and 320 µm $\pm$ 16 µm respectively. Figure 3 a,c) is an earlier version of this set up where 3D printed mounts were used to align the 200mW laser to the multimode collimator which is directly connected to the fibre optic feedthrough flange. The 800mW laser diode was enclosed in a 30mm x 30mm x 66.5mm rectangular housing (which contained a fan) and its power could be varied using pulsed width modulation (PWM). It was placed in a bespoke 3D printed holder to accommodate its larger size and aligned with the multimode collimator on a kinematic mount using the same light injection set up as Figure 2b).

The photoelectric effect will only occur if the photon energy is greater than the workfunction of the LaB$_6$ crystal. Unfortunately, if the surface of the crystal becomes contaminated the workfunction of the crystal can increase. This is not an issue in ultra-high vacuum systems, however the vacuum pressure of the gun chamber used was relatively poor ($1 \times 10^{-6}$mbar). To prevent unwanted contamination the crystal was cleaned before each of the following experiments by heating it to ~1443K with a resistive current of 1.4A for at least twenty minutes, as per the manufacturer's instructions. Using the LaB$_6$ Kimball Physics specification sheet, the temperature of the LaB$_6$ crystal could be deduced from the applied resistive current. It was found that if the crystal was at a temperature below ~904K (~0.8A resistive current) the photoelectric effect could not be observed implying the contamination of the crystal occurred at and below this crystal temperature. To prevent this unwanted contamination, during normal operation the crystal was heated, and the indicated temperatures are stated in each experiment.

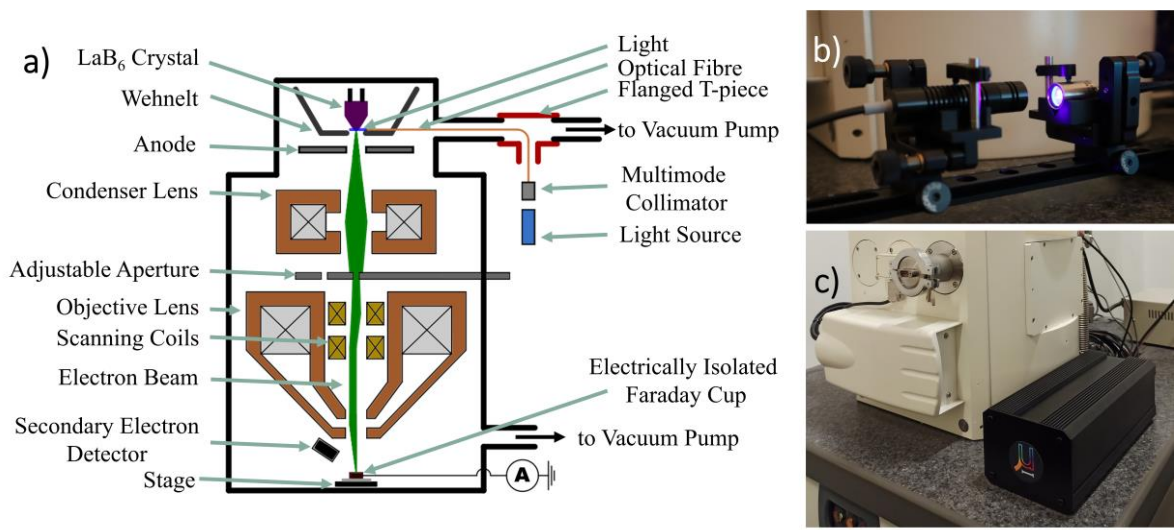

*Figure 2 a) A schematic of the photoelectron system adapted onto the ZEISS EVO SEM, b) highlights the laser diode coupled to the interior fibre optics through a multimode collimator and an optical fibre and c) the light injection set up in its light tight box.*

## Results and Discussion

A key test in our experimental work was to verify that photoelectrons, not thermionic electrons were being produced from the emitter when struck by the laser light. For this experiment the LaB$_6$ crystal was heated to ~1443K (via a 1.4A resistive current) for twenty minutes to remove contamination from the surface. Then it was left at a resistive current of 0.96A (LaB$_6$ temperature of ~1064K) for 20 minutes. A laser diode of wavelength 405nm and power of 200mW was then used to take a SEM image of titanium powder (often used in metal 3D printing) mounted on a carbon tab. This is shown in Figure 3a), which is an earlier set up of the light injection system shown in Figure 2b). Halfway through the SEM image the laser was turned off and we see the intensity of the SEM image immediately decreases to a negligible intensity. This confirms that the electrons were being produced from the light striking the $LaB_6$ crystal.

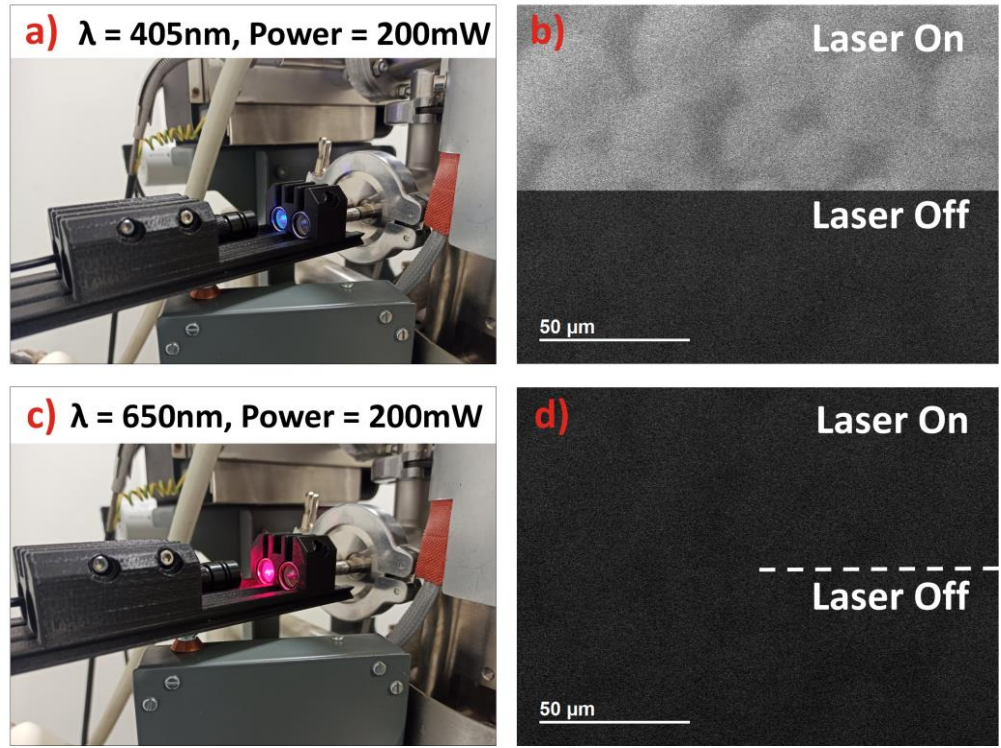

*Figure 3 a) The exterior of the set up with a laser of power 200mW and wavelength 405nm, b) SEM image of titanium powder on a carbon tab where a laser of wavelength 405nm was turned on for the top half of the image and then turned off for the bottom half of the image. c) is the same set up as a) with a laser of the same power but with a wavelength of 650nm, d) is the same laser on/off experiment as b) except a laser of wavelength 650nm was used. Please note an aperture of 750 μm was in the SEM, the $LaB_6$ crystal was at a temperature of ~1064K (had a resistive current of 0.9A through it) and EHT=15kV for the duration of both these experiments. The brightness and contrast were kept the same for both images in b) and d).*

The experiment was then repeated with a laser of wavelength 650nm with an identical power of 200mW. This laser (at 650nm / 1.91eV) emits photons with insufficient energy to yield photoelectrons. A SEM image was then taken with this laser on/off again (Figure 3d)). It is clear from Figure 3d) that there is no change in intensity when the laser is on or off and the intensity remains relatively negligible independent of the state of the laser. This confirms that the electrons being produced by the UV laser are photoelectrons not thermionic electrons being produced from the laser wattage heating up the $LaB_6$ crystal. This in turn, verifies that the intensity seen with the 405nm laser was photoemission.

As we are unable to introduce a faraday cup midway along the SEM column the current produced at the tip of the $LaB_6$ thermionically or via the photoelectric effect could not be experimentally measured. We were however able to estimate the photoelectron current ($I_{ph}$) and thermionic current ($I_{th}$) to strike the sample. This was achieved by disconnecting the stage specimen current monitor (i.e., electrically isolating the stage) and measuring the current $I_{ph}$ at the SEM stage using a Faraday cup connected to a picoammeter. To minimise measurement uncertainty the largest possible aperture of 750 µm was placed in the SEM for the following experiments. The current when the laser (405nm and power 400mW) was off was taken as $I_{th}$ and the current when the laser was on minus $I_{th}$ was taken as $I_{ph}$.

Using the LaB$_6$ Kimball Physics specification sheet, the thermal current at the microflat at the tip of the LaB$_6$ could be deduced from the applied resistive current for different LaB$_6$ temperatures. This meant the ratio of the thermionic current at the microflat for LaB$_6$ temperatures between ~1282K-1443K (1.2A-1.4A) versus the thermionic current at the stage ($I_{th}$) can be determined during normal thermal emission. For example, at 1282K the thermionic emission at the tip was determined to be ~30 µA and the current detected by the Faraday cup $I_{th}$ was ~4.5nA, so 0.02% of the current made it down the column to the sample. From the calculation of this ratio between ~1282K-1443K, we inferred only ~0.02-0.1% of the thermionic current generated at the microflat of the crystal propagates down the column from the tip. Please note this value does not consider any thermionic electrons generated outside the microflat of the crystal, however the bias of the wehnelt has been set to try optimally select electron from the microflat and supress those electrons emitted elsewhere from the crystal.

Using this thermal ratio and the $I_{ph}$ measured at the sample the number of photoelectrons generated at the tip of the LaB$_6$ could be calculated. At a LaB$_6$ temperature of 1282K, $I_{ph}$ at the sample was measured as ~0.84nA therefore the photocurrent emitted from the LaB$_6$ tip was calculated to be ~5µA from the thermal ratio calculations. Using equation 3 it was calculated that 24 µA of photocurrent could be emitted at the LaB$_6$ tip if 100% of the laser light made it from the laser to the LaB$_6$ tip. As with the thermionic emission mode, the bias of the wehnelt is optimally selecting electrons emitted from the microflat of the crystal. This suggests that approximately 23% of the laser light strikes the tip of the LaB$_6$ crystal and causes photoelectrons to be emitted. When calculated for temperatures ~1282K-1443K it was found this value was between 23%-82%. We believe this number varies with temperature as it is a product of the efficiency of light to strike the crystal and the efficiency of the photoelectron emission from the crystal. This number indicates that there is large room for design improvement in our implementation. This could include parabolic mirrors or ball lenses to improve the focusing of the laser, as in the current set up the fibres are simply pointed at the crystal and the large angle at which the light exits the fibre ensures some photons will strike the tip of the crystal. Some of the light lost could also be due to laser losses such as reflectivity at the LaB$_6$ surface. The calculation of the values mentioned can be found in more depth in the supplemental information.

Using PWM on an 800mW 405nm laser, the duty cycle could be changed to decrease the power of the light striking the crystal. There is a linear relationship between duty cycle and output power of a laser during PWM (Horowitz & Hill, 2015). We measured I$_{ph}$ and I$_{th}$ at different laser powers using decreasing duty cycles. This is shown in Figure 4a) where we see I$_{th}$ remains the same independent of the duty cycle however the I$_{ph}$ increases with an increasing duty cycle. From observing Figure 4a), there appears to be some minimum duty cycle at which the photoelectrons are stimulated. There is also a curve in the I$_{ph}$ which may be the result of a multiphoton effect. In future we intend to explore the variability of the laser power quoted from the manufacturer to determine how the duty cycle and output power directly correlates to the photoelectron current detected.

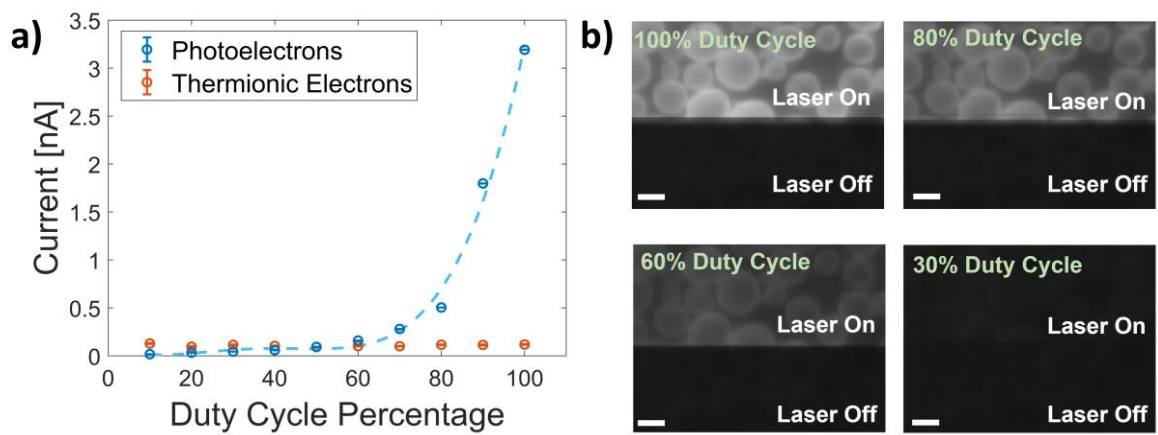

*Figure 4 a) Trend of beam current detected versus the PWM duty cycle of an 800mW laser. b) SEM images of titanium powder on a carbon tab when the laser was on/off at different duty cycles which corelates with the duty cycles indicated in a). Note the temperature of the LaB$_6$ was ~1264K, EHT=15kV and an aperture of 750μm was in the SEM when a) and b) were recorded. The scale bar is 20μm and the brightness and contrast were kept the same for each image in b).*

Figure 4b) corresponds to images taken at the same experimental conditions at four of the different duty cycles in Figure 4Figure 1a) (the rest of the different duty cycle images can be found in the supplemental Figure 4). Due to the extremely large aperture placed in the SEM, we note that the resolution of the images in Figure 4b) are not very high. Images at the different duty cycles with an aperture of 30 µm in the SEM can be found in Figure 5 in the supplemental information and with the subsequent reduction of spherical aberration the resolution drastically improves.

When the smaller aperture of 30 µm was placed in the SEM the max photoelectron current produced at a duty cycle of 100% was reduced from ~3.19nA to ~0.12nA. Operating the SEM in normal thermionic mode with a 30 µm aperture inserted, we found ~3.00nA of current was detected using the specimen current monitor. This will be a slight underestimation of the current due to some backscatter electrons not being recorded by the specimen current monitor. There is a clear disparity between the operating current with the aperture in during thermionic and photoelectron mode. While the level of photocurrent produced was high enough to form images at slow scan speeds with the 30 µm aperture inserted, it is clear from these experiments that the current design of the photoelectron emitter is producing sub optimal photocurrent for everyday operation of the SEM which may require even smaller apertures.

We will design further experiments to undertake a more quantitative study of the retrofittable photoelectron emitter, this will involve seeking to improve alignment, reproducibility, and durability of the system. We hope that in a future system, we can work at a better vacuum condition and lower $LaB_6$ temperature. This will lead to a more predictable value of the work function and an overall increase in photoelectron current which will improve the ratio of photoelectrons produced relative to thermionic electrons.

## Conclusion

Using the photoelectric effect, optical components, and a $LaB_6$ cathode we have demonstrated a potential means of how to produce a low energy spread retrofittable electron gun. Despite only a fraction of the initial laser light striking the cathode, SEM images were formed with the emitter, and it was confirmed that this was achieved via photoelectrons. Using PWM on the laser, it was found that photoelectron current increased with increasing laser duty cycle while thermionic current remained the same. The photocurrent produced however was significantly less than that of a thermionic $LaB_6$ gun during normal operation. This initial proof of concept design proved the physics behind the photoelectron emitter is sound and shed light on the different engineering directions that could greatly improve the current emitter design. This includes a more accurate focusing system to optimise the light injection set up and a better vacuum condition to reduce the contamination of the crystal. Once these challenges are addressed, the photoelectron current emitted should greatly increase. This increase in current combined with the potential low energy spread of the source will make it an ideal candidate electron emitter for improved low voltage SEM imaging.

## Acknowledgments

The authors acknowledge the support of the Advanced Microscopy Laboratory of the Centre for Research on Adaptive Nanostructures and Nanodevices (CRANN). FQ is financially supported by the Provost's Project Award and LJ is supported by an SFI/Royal Society Fellowship (grant number URF/RI/191637).

## Conflict of interest

Conflict of interest: The authors declare none.

# Supplemental Information for: A Retrofittable Photoelectron Gun Proposal for Improved Low Voltage Imaging

*Frances Quigley[1,2*], Clive Downing[2], Cormac McGuinness[1], Lewys Jones[1,2]*

*1. School of Physics, Trinity College Dublin, Dublin 2, Ireland*

*2. Advanced Microscopy Laboratory, Centre for Research on Adaptive Nanostructures & Nanodevices (CRANN), Dublin 2, Ireland*

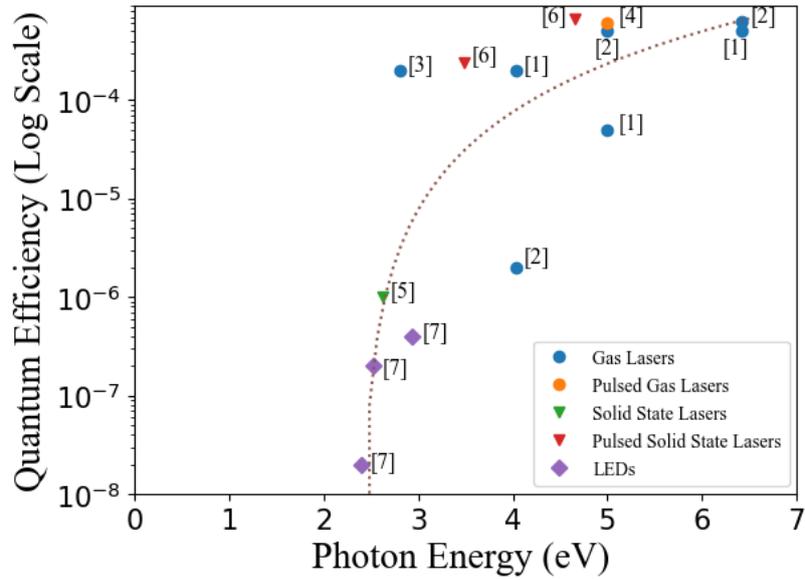

*Figure 5:Literature values of photon energy versus quantum efficiency for LaB$_6$ where the following references were used; [1](Oettinger, 1990), [2](Lafferty, 1951), [3](May et al., 1990), [4](Qian et al., 1995),[5](Sawa et al., 2017),[6] (Leblond et al., 1996), ,[7](Konishi et al., 2012).*

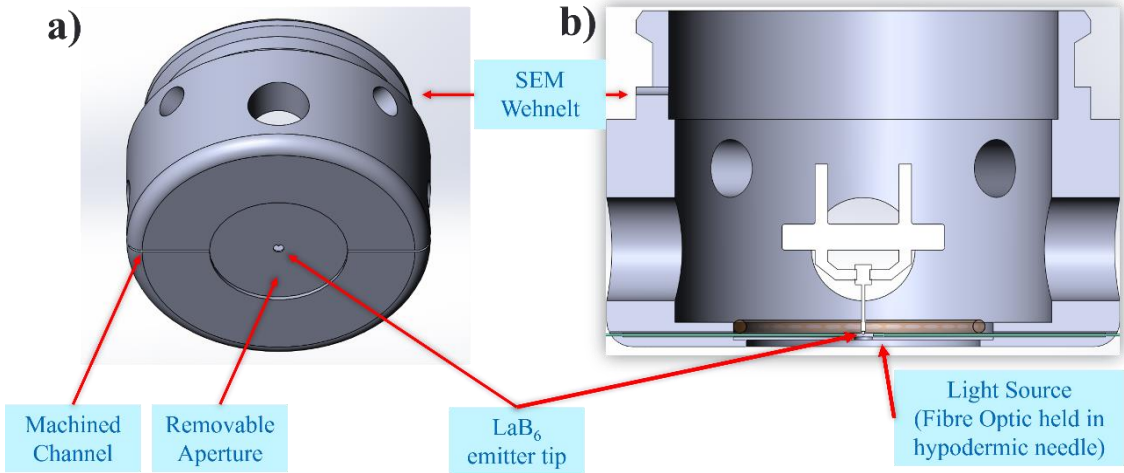

*Figure 6: a) Outside view of ZEISS EVO SEM adapted wehnelt with machined channel for hypodermic needle to be placed in. b) Interior of adapted wehnelt highlighting the optical fibre fed through the hypodermic needle such that the light will strike the LaB₆ tip.*

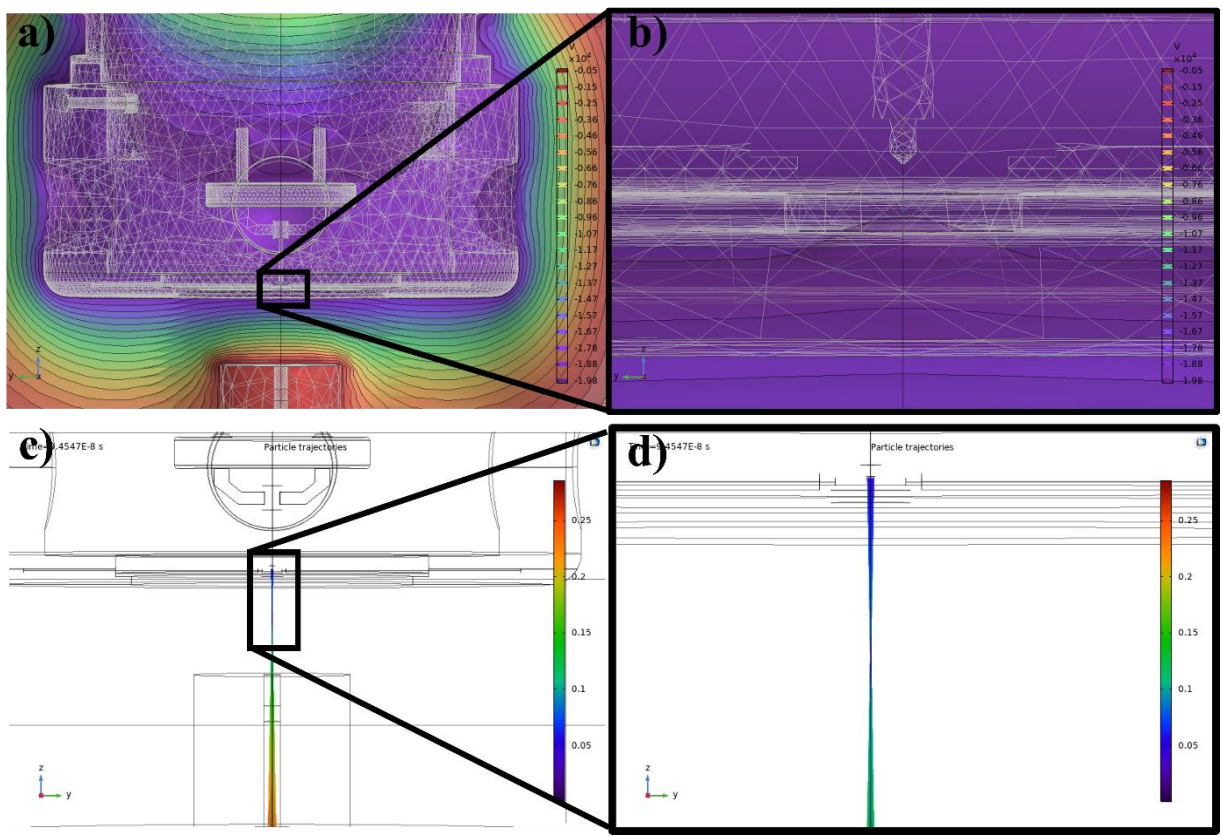

*Figure 7: a), b) COMSOL electrostatic simulation of the electric field applied to the adapted wehnelt with needles. b), c) The particle trajectories of the emitted electrons. Note the colour axis in c) and d) indicates how radially far away the electrons are from the optic axis and is given in mm.*

# Calculating the ratio of the current at the LaB$_6$ tip and the current detected at the sample

The ratio of the thermionic current at the LaB$_6$ tip and the thermionic current detected at the sample stage needed to be calculated. The first step was to calculate the current emitted at the LaB$_6$ at a certain temperature. Our crystal was a Kimball Physics LaB$_6$ cathode, model ES-423E, style 90-20. This means it had a 20 µm diameter microflat at its tip. Kimball physics has released a LaB$_6$ specification sheet which has a graph of emission current versus the heating current of the crystal for 15,40,100,200 µm diameter microflats. The graph refers to emission from only the microflat of the diameter specified in the legend for each plot. Our emission current would therefore be slightly above the 15 µm diameter microflat. We will take our emission values off the 15 µm microflat graph but note that our ratio is most likely going to be slightly larger than expected due to this approximation. Heating current versus true temperature of the LaB$_6$ crystal is also given in the specification sheet so the temperature of the crystal versus the emission current can also be calculated. The thermionic current at the sample I$_{th}$ can be measured by electrically isolating the stage and measuring the current detected by a faraday cup on the stage which is connected to a picoameter. We had the largest aperture of 750 µm in for these measurements. The ratio of thermionic current at the sample dived by the thermionic current at the tip can then be calculated. For example, at 1282K the thermionic emission at the tip was determined to be ~30 µA, the current detected by the Faraday cup was ~4.5nA, so 0.02% of the current made it down the column to the sample. From 1282K-1443K the percentage of current that travelled down the column varied from 0.02%-0.1%.

The photoelectron current at the sample could also be similarly measured using the Faraday cup. By turning the laser off and on and measuring the difference in the current detected at the sample the photoelectron current at the sample I$_{ph}$ could be measured. By dividing I$_{ph}$ by the ratio of the thermionic current to make it down the column at the same LaB$_6$ temperature we can therefore calculate the photocurrent emitted at the tip. At the previous LaB$_6$ temperature of 1282K I$_{ph}$ detected was 0.8nA so the calculated photoemission at the tip would be ~ 5µA.

Using equation 3 in the paper, we can calculate the total photocurrent produced if 100% of the light emitted by the laser were to strike the crystal. The quantum efficiency in this graph is taken from the trendline in the supplemental Figure 1. From equation 3 it was calculated that theoretically 24µA of photocurrent should be produced from a 400mW 405nm laser diode. This is clearly an overestimation and implies only 23% of laser light gets from the diode through the optical components to strike the LaB$_6$ tip and cause photoemission at a crystal temperature of 1282K. For crystal temperatures ~1282K-1443K it was found this value was between 23%-82%.

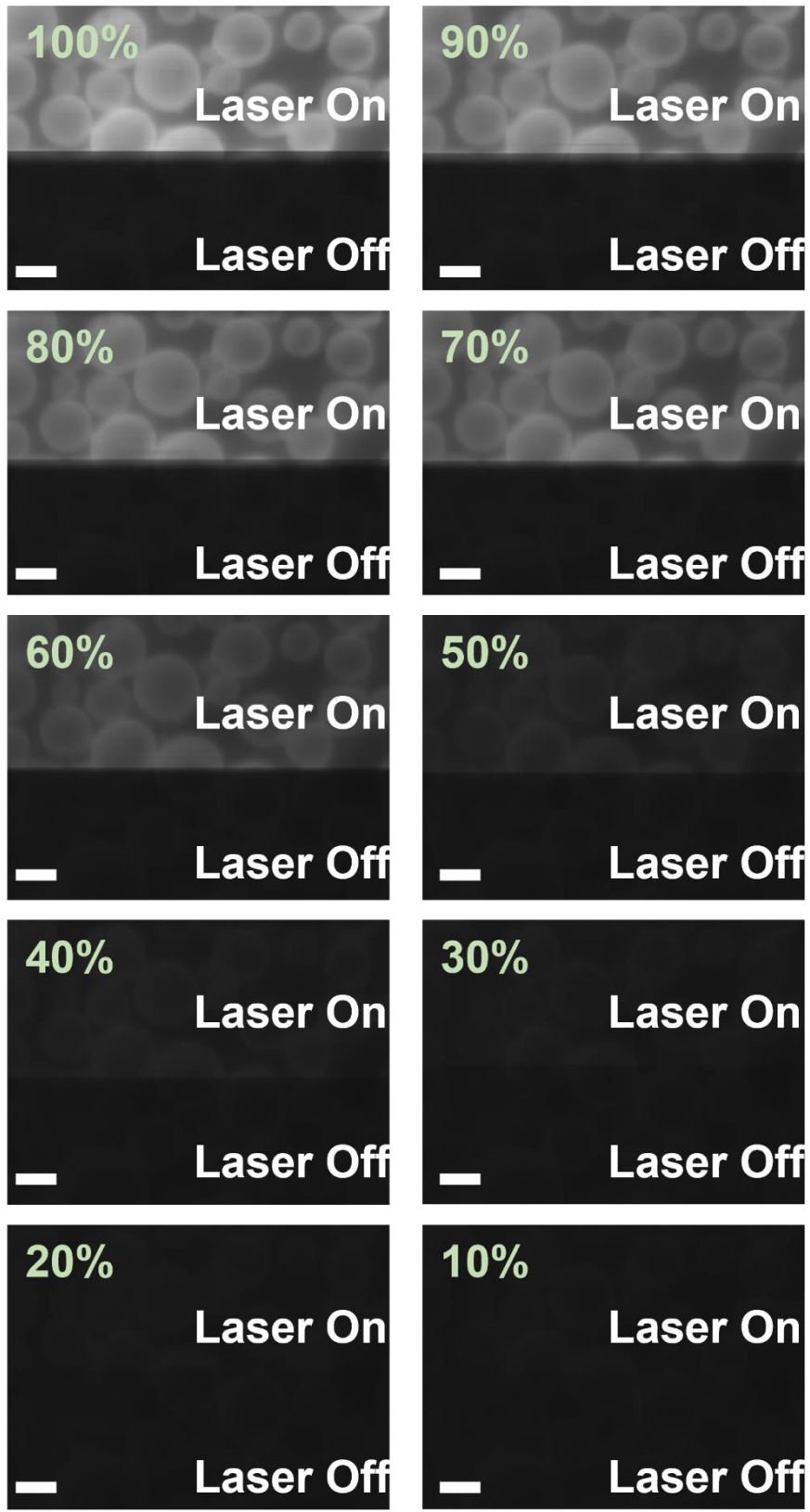

*Figure 8: SEM images taken when the laser was on/off at different duty cycles percentages which are indicated on the top left corner of the image. Note the temperature of the $LaB_6$ was ~1264K, EHT=15kV and an aperture of 750μm was in the SEM when these images were taken. The scale bar is 20μm and the brightness and contrast were kept the same for each image.*

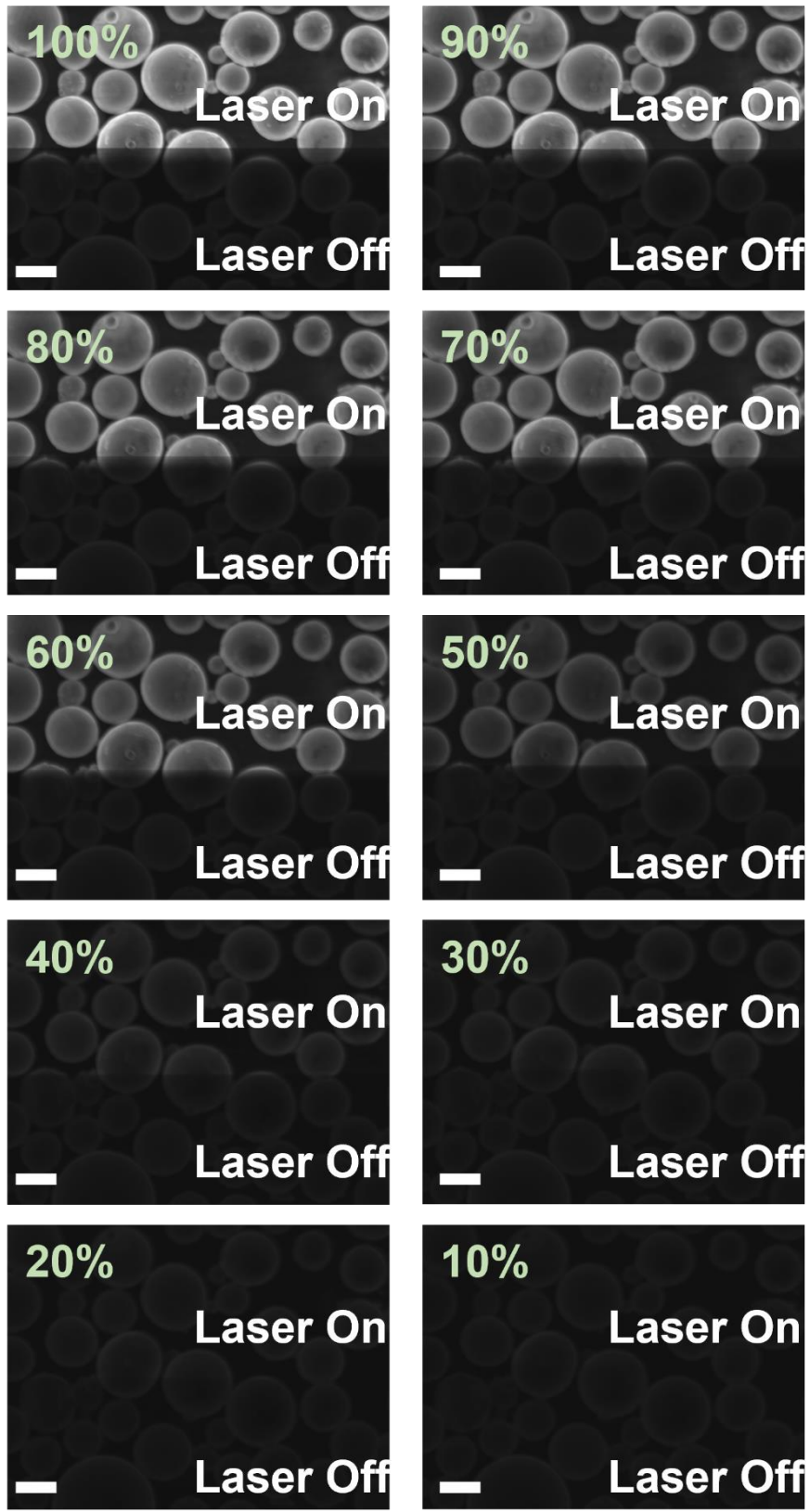

*Figure 9: SEM images taken when the laser was on/off at different duty cycles percentages which are indicated on the top left corner of the image. Note the temperature of the LaB₆ was ~1264K, EHT=15kV and an aperture of 750μm was in the SEM when these images were taken. The scale bar is 20μm and the brightness and contrast were kept the same for each image.*